\documentstyle[12pt]{article}

\newcommand{\be}{\begin{equation}}
\newcommand{\ee}{\end{equation}}
\begin{document}
\begin{flushright}
\vspace{-1mm}
 FIAN/TD/15--95\\
\vspace{-1mm}
{September 1995}\\
\end{flushright}\vspace{2cm}
\begin{center}
{\large\bf
HIGHER-SPIN GAUGE INTERACTIONS FOR MATTER FIELDS IN TWO DIMENSIONS}
\vglue 1  true cm

{\bf M.A.~VASILIEV}

\medskip
I.E.Tamm Department of Theoretical Physics, Lebedev Physical Institute,\\
Leninsky prospect 53, 117924, Moscow, Russia
\medskip
\end{center}

\begin{abstract}
\baselineskip .4 true cm
\noindent
\end{abstract}

\section{Introduction}
Many of important properties of
integrable systems and conformal models in two dimensions
originate from underlying infinite-dimensional symmetries.
Higher-spin (HS) extensions of the Virasoro symmetry
acquired much attention during recent years.
In particular, d2 conformal matter models with gauged HS
symmetries have been extensively studied for
the cases of $W_N$ \cite{fin}, $\omega_\infty $ \cite{win} and
$W_{1+\infty}$ \cite{BB,WIN} algebras.
A d=2 model for pure  gauge $W_{1+\infty}\,$ HS
fields was proposed in \cite{FL}.
The $W_{1+\infty}$ algebra investigated originally
in \cite{mat} is getting now a
wide area of applicability including, e.g.,
fractional Hall effect
\cite{hall} and
KP hierarchy\footnote{As it often happens in the literature, we use
the same name $W_{1+\infty}$ both for the full algebra that
admits a non-trivial central extension and for its centerless part
which in turn is equivalent to the Moyal bracket \cite{MOY}.}
\cite{KP}. In the context
of field theory $W_{1+\infty}$ - type symmetries were originally introduced as
HS symmetries in \cite{BB} for the case of d=2, and in \cite{W,OP} for
d=4. The name $W_{1+\infty}$ was suggested in \cite{name} in the context of
the analysis of possible Lie algebra extensions of the Virasoro algebra.

In refs. \cite{more} and \cite{un,d3} consistent dynamics of HS
gauge fields interacting among themselves and with the lower-spin matter fields
in d=4 and d=3 was formulated in terms of gauge fields corresponding to the
appropriate versions of $W_{1+\infty}$-type algebras.
An interesting problem that remained unsolved for some years
was to apply the methods developed for d=3 and d=4 HS problems
to d2 models.
The goal of this letter is just to announce a new model which describes HS
gauge interactions of boson and fermion matter fields
in d=2 along the lines of the approach developed in [13-15].
As expected, the d2
HS dynamics turns out to be much simpler than that in d=3 and 4.
However, the
formalism in d=2 has a number of specific properties
and does not amount to a straightforward reduction of
the higher-dimensional models. In particular, the full set of gauge fields,
which is shown below to describe d2 HS-matter interactions,
differs from that proposed in \cite{BB,FL}, rather corresponding to
a current extension of the $W_{1+\infty}$ algebra.

Being a counterpart of the $W_{1+\infty}$ gauge models
discussed in
\cite{BB,WIN} in the sense that it is based on bilinear HS currents
involving higher derivatives, the presented model has a number of
physically important distinctions from the models of
\cite{BB,WIN}. In particular, no vanishing current
constraints on the matter fields are present in our model.
The model is formulated in an explicitly HS gauge invariant
and general coordinate invariant fashion. A natural d2 background
is anti-de Sitter (AdS) space-time.

The proposed model is not conformal. Instead,
the full equations of motion in this model have a form of some
zero-curvature equations and covariant constantness conditions without
any additional constraints so that the model turns out to be
integrable. This unexpected property is specific for d=2
and allows one to write down a simple action principle for the model.

\section{Basic Algebraic Structures}

Analogously to the construction developed previously for
HS theories in d=3 and 4 (see [10,11,13-15] and
references therein) the basic algebraic structure is the associative algebra
$A$ of power series in the generating elements $\hat{y}_\pm$ and
$\hat{z}_\pm$ obeying the commutation relations
\be
\label{ha}
[\hat{y}_- ,\hat{y}_+ ] = -2i\,,\qquad [\hat{z}_- ,\hat{z}_+ ] = 2i\,,\qquad
[\hat{y}_\alpha ,\hat{z}_\beta ] = 0\,\qquad \alpha ,\,\beta =\pm\,.
\ee
Any $\hat{a}\in A$ can be cast into the form
\be
\label{el}
\hat{a}=\sum_{n,m=0}^\infty a^{\alpha_1 \ldots \alpha_n\,\beta_1
\ldots \beta_m}
\hat{y}_{\alpha_1} \ldots \hat{y}_{\alpha_n}
\hat{z}_{\beta_1} \ldots \hat{z}_{\beta_m}\,,
\ee
where the coefficients
$ a^{\alpha_1 \ldots \alpha_n\,\beta_1 \ldots \beta_m}$ are
totally symmetric in $\alpha$ and $\beta$ that implies  Weyl
ordering of the operators $\hat{y}_{\alpha}$ and $\hat{z}_{\beta}$.
It is also convenient to use the following equivalent set of variables
\be
\label{uv}
\hat{u}_\alpha =\frac{1}{2}\,(\hat{z}_\alpha -\hat{y}_\alpha )\,, \quad
\hat{v}_\alpha =\frac{1}{2}\,(\hat{y}_\alpha +\hat{z}_\alpha )\,,\quad
[\hat{v}_\pm ,\hat{u}_\mp ] =\mp i \,,\quad [\hat{v}_\alpha ,\hat{v}_\beta ] =
[\hat{u}_\alpha ,\hat{u}_\beta ]=0\,.
\ee

Practically, to work with the algebra $A$ it is most useful to use
the technics of symbols of operators \cite{sym,OP}.
Namely, given element $\hat{a}$ (\ref{el})
one introduces its symbol of the form

\be
\label{els}
a=\sum_{n,m=0}^\infty a^{\alpha_1 \ldots \alpha_n\,\beta_1 \ldots \beta_m}
y_{\alpha_1} \ldots y_{\alpha_n}
z_{\beta_1} \ldots z_{\beta_m}\,,
\ee
where
$y_{\alpha}$ and $z_{\beta}$ are commuting variables. By definition,
a star product $(a*b)$ is the symbol of the operator $\hat{a}\hat{b}$.
One can use the following useful formula for the Weyl symbol star product
\begin{equation}
\label{weyl}
(f* g)(z,y)=(2\pi )^{-4} \int d^2 s d^2 t d^2 p d^2 q\, f(z+s,y+p) g(z-t,y+q)
exp[i(s_\alpha t^\alpha +p_\alpha q^\alpha )]\,.
\end{equation}

An important property of $A$ is that it contains an element $\Pi $
which possesses properties of the projection operator, $\Pi^2 =\Pi$, and
behaves
as the vacuum vector for the operators $\hat{u}_\pm$ and $\hat{v}_\pm$, i.e.
$\hat{v}_\pm \Pi =0$ and $\Pi \hat{u}_\pm  =0$.  It has the following explicit
realization in terms of Weyl symbols

\be
\label{fock}
\Pi=\frac{1}{4}\, exp(i z_\alpha y^\alpha )\,.
\ee

The algebra $A$ is $Z_2$ - graded with odd (even) power polynomials
in (\ref{el}) considered as odd (even) elements of $A$. This
is just the ordinary boson-fermion grading since the indices
$\alpha$ and $\beta$ are interpreted as spinor indices in field
theoretical applications. The algebra $A$
admits the following unique supertrace operation \cite{OP}
\be
\label{strA}
str_A (\hat{a})=a_0\,,
\ee
where $a_0$ is a constant term in the expansion (\ref{el}).
This supertrace operation is a linear mapping obeying the general property
\be
str(\hat{a}\hat{b})=(-1)^{\pi (\hat{a})\pi (\hat{b})}
str(\hat{b}\hat{a})\,,
\ee
where $\pi (\hat{a})$= $0$ or $1$ is the parity of $\hat{a}$.
Note that this property allows one to construct invariant
polylinear forms with the aid of the supertrace in the standard way as
$str(\hat{a_1}\hat{a_2}\ldots\hat{a_n})$.

To formulate the d2 higher-spin dynamics
it is useful to extend the algebra $A$ to ${\cal A}$ in the following
general way.
Given associative algebra $A$ and some projection operator $\Pi\in A$,
one defines the
algebra ${\cal A}$ such that its general element ${\bf a}\in {\cal A}$ is
equivalent to a set
of four elements of $A$,
${\bf a}$=$\{a,|a\rangle ,\langle a|,\langle a\rangle \}$,
obeying the properties
\be
\label{pr}
\{{\bf a}\in {\cal A}\vert\,\,\,
a,|a\rangle ,\langle a|,\langle a\rangle \in A\,;
\,\,|a\rangle \Pi =|a\rangle ,\,
\Pi\langle a| =\langle a| ,\,
\langle a \rangle \Pi = \Pi \langle a \rangle =\langle a \rangle\}\,.
\ee
The product law  $\circ$
in ${\cal A}$ is defined via the product law in $A$ as follows
\be
\label{fp}
{\bf a}\circ {\bf b} =\{ab + |a\rangle\langle b|\,, a |b\rangle
+|a\rangle \langle b \rangle \,, \langle a| b+\langle a \rangle\langle
b|\,,\langle a||b\rangle + \langle a \rangle\langle b \rangle \}\,.
\ee
This product law is associative. Note that supertrace operation
$str_A$ in
$A$ induces the supertrace operation $str_{\cal A}$ in ${\cal A}$
\be
\label{str}
str_{\cal A} ({\bf a})=str_A (a+\langle a\rangle )\,.
\ee

We will use this construction with the projection operator (\ref{fock})
to embed
all matter and auxiliary fields into the adjoint representation
of ${\cal A}$.

\section{Full Nonlinear Dynamics and Vacuum Solution}

In this section we summarize the final results for the
full non-linear equations and the action principle of the
model and give a vacuum solution which
is shown in the subsequent sections to
lead to a proper linearized dynamics.

To describe the HS gauge interactions of d2 matter
fields
($x^\nu$
are space-time coordinates ($\nu =0,1$), $dx^\nu$ are
anticommuting differentials,
$d=dx^\nu \frac{\partial}{\partial x^\nu}$)
we introduce the gauge one-form
${\bf W}(x|z_\alpha ,y_\alpha )$=
$dx^\nu{\bf W}_\nu (x|z_\alpha ,y_\alpha )$,
and the matter field zero-form ${\bf B}(x|z_\alpha ,y_\alpha )$
in the adjoint representation of ${\cal A}$, i.e.
\be
{\bf W}
=\{W,\,|W\rangle ,\, \langle W| ,\,
 \langle W \rangle \},\qquad
{\bf B}
=\{B,\,|B\rangle ,\, \langle B| ,\,
 \langle B \rangle \}\,.
\ee

The full system of equations for interacting d2 matter fields has
a simple form of zero-curvature conditions:
\be
\label{1f}
{\bf R}\equiv d{\bf W} + {\bf W}\circ\wedge {\bf W} =0\,,\qquad
d{\bf B} + {\bf W}\circ
{\bf B}-{\bf B}\circ{\bf W} =0\,.
\ee
These equations can be derived from the B-F type action principle
\be
\label{action}
S= \int_{M_2} str_{\cal A}({\bf B}\,{\bf R})\,.
\ee

The model becomes dynamically non-trivial because
the 0-form
${\bf B}$ is supposed to have a nonvanishing vacuum value
of the form\footnote{Note
that the physical vacuum values of the fields ${\bf B}$ and ${\bf W}$ have
nothing to do with the vacuum $\Pi $ of the algebra $A$ of auxiliary spinor
variables.}
\be \label{vacb} {\bf B}_{vac} =
\{N,0,0,0\}\,,\qquad N={1\over 4i} \{\hat{z}_- ,\hat{z}_+\}\,.  \ee A physical
vacuum value of the gauge 1-form ${\bf W}$ is \be \label{vacw} {\bf W}_{vac} =
\{\omega^{gr},0,0,0\}\,,\qquad \omega^{gr} (x) = h^{+} (x) L^+ +h^{-} (x) L^- +
\omega (x) L^0\,,
\ee
where
\be
\label{gen} L^\pm = \frac{i}{4} (\hat{y}_\pm )^2\,,\qquad   L^0 = \frac{i}{4}
\{\hat{y}_+ ,\hat{y}_- \}
\ee
 obey the $sl_2$ commutation relations,
\be
\label{com}
[L^0 , L^\pm
] = \pm 2 L^\pm\,, \qquad [L^- ,L^+ ]=L^0 \,.
\ee

 One-forms $h^\pm (x)=dx^\nu h^\pm_\nu (x)$ and $\omega (x)=dx^\nu \omega_\nu
(x) $ describe inverse zweibein and Lorentz connection, respectively.  The
components of the gravitational field-strength two-form,

\be
R^{gr} = d\omega^{gr} +\omega^{gr}\wedge \omega^{gr}
= R^{+} L^+ +R^{-} L^- + R^0 L^0
\ee
identify, respectively, with the torsion tensor, $R^{+}$, $R^{-}$, and with the
Riemann tensor, $R^0$, shifted by a cosmological term $h^- \wedge h^+$.
The vacuum gravitational field is supposed to obey the
zero-curvature conditions \be \label{0c}
R^0 =d\omega +h^- \wedge h^+ =0\,,\qquad R^\pm =dh^\pm \pm 2\omega \wedge h^\pm
=0\,.
\ee

Under
the condition that the zweibein $h_\nu^\pm$ is invertible these conditions
imply that the Lorentz connection expresses in terms of
$h_\nu^\pm$ by virtue
of the metric postulate while $h_\nu^\pm$ describes the two-dimensional
AdS space. On the other hand, (\ref{0c})
along with the fact that bilinears in the oscillators $\hat{y}$
form a subalgebra $sl_2\subset A\subset {\cal A}$ implies
that the first of the equations (\ref{1f}) is satisfied. The second
one is also true because the vacuum value $N$ of ${\bf B}$ depends only
on  $\hat{z}$ and therefore commutes with the background gravitational
field (\ref{vacw}) due to (1).

HS gauge fields correspond to higher-order terms of the
expansion of
${\bf W}(x|z_\alpha ,y_\alpha )$ in powers of the auxiliary spinor
variables.
The gauge connection ${\bf W}$ and the matter field ${\bf B}$ have
the standard transformation laws under the
HS gauge transformations with the parameter
${\bf \xi} (x|z_\alpha ,y_\beta )$,
\be
\label{g1}
\delta {\bf W} = d{\bf \xi} +{\bf W}\circ {\bf \xi}-{\bf \xi}\circ
{\bf W}\,,\qquad
\delta {\bf B} = {\bf B}\circ {\bf \xi}- {\bf \xi}\circ {\bf B}\,,
\ee
which leave invariant the equations (\ref{1f}) and the action (\ref{action}).
Because of using the exterior algebra formalism,
general coordinate invariance is explicit too.

Let us note that
a part of the symmetry that acts
linearly on physical states consists of the subalgebra spanned by
elements commuting with ${\bf B}_{vac}$. The gauge parameters of this
subalgebra are of the form
\be
{\bf \xi}=(\xi_{vac}\,,0,0,\langle\xi_{vac}\rangle )
\ee
with an arbitrary Abelian parameter
$\langle\xi_{vac}\rangle$ and the parameter $\xi_{vac}$
of the form
\be
\label{vs}
\xi_{vac}=\sum_{n,m,k=0}^\infty \xi_{n,m,k} (x) N^k
(\hat{y}_- )^n \,(\hat{y}_+ )^m\,.
\ee
Since $N$ (\ref{vacb}) commutes with the oscillators $\hat{y}_\pm$, the
generating elements of the $W_{1+\infty}$ algebra, one is left with the
non-negative part
of the loop extension  $\tilde{W}_{1+\infty}$  of $W_{1+\infty}$.

The purely topological form of the
action (\ref{action}) is analogous to the
topological form of the d2 gravitational action
discussed in \cite{gr} and to the HS action proposed in \cite{FL}.
This analogy is not exact however because in the latter
models the zero-curvature equations are true
in absence of matter and do not
describe propagating degrees of freedom while the equations (\ref{1f}) are
shown below to describe interactions of propagating scalar and spinor fields.
As is demonstrated in the next section,
this is possible because of using infinite
multiplets of fields.

Another important point is that the non-vanishing vacuum value
of the zero-form ${\bf B}$ (\ref{vacb}) breaks down spontaneously
 the antisymmetry
of the action $S$ (\ref{action}) under the transformation
${\bf B} \rightarrow$$-{\bf B}$.
The vacuum value
(\ref{vacb}) leads effectively to some
$W^2$ - type terms in the action that opens a way to a proper diagonalization
of
the action at the linearized level. Practically, a problem of reducing the
quadratic part of the action (\ref{action}) to the standard form is
highly involved due to presence of an infinite set of
auxiliary fields and will be considered elsewhere.

\section{Free Matter Equations in d2 AdS Space}

In this section we reformulate free equations for matter fields
in d2 AdS space described by the zero-curvature conditions (\ref{0c})
in the form of some covariant constantness conditions along the lines of
 the general ``unfolded formulation'' approach proposed in \cite{un}.
Importance of this reformulation for the analysis below is due to the
fact that it is this form to which the full nonlinear equations
reduce in the linearized approximation in the sector of matter fields.

Consider the following system of equations
\be
\label{Cn}
D\phi_n =\alpha (n) h^- \phi_{n+2} +\beta (n) h^+ \phi_{n-2}\,,
\ee
where $D$ is the Lorentz covariant derivative,
\be
D\phi_n =d\phi_n +n \omega \phi_n\,.
\ee
This system is formally consistent (i.e. the Bianchi
identities are satisfied) provided that
the numerical parameters
$\alpha (n)$ and $\beta (n)$ obey the condition
\be
\label{ab}
\alpha (n)\beta (n+2) = \mu +1/4\, n(n+2)\,
\ee
and zero curvature conditions (\ref{0c}), describing the vacuum
AdS geometry, are satisfied. Here $\mu$ is an arbitrary numerical
parameter. Note that the ambiguity in the coefficients $\alpha (n)$
and $\beta (n)$, which is not fixed from (\ref{ab}), is
irrelevant and reflects a
freedom in the rescaling  $\phi_n\rightarrow \gamma (n)\phi_n$.

Now we observe that the infinite system of equations (\ref{Cn})
is equivalent to the dynamical equations of free d2 fields
of arbitrary mass $m^2 =\mu$. Here boson and fermion fields are described
by the set of the fields $\phi_n$ with $n$ even and odd, respectively.

To make sure that, e.g., the equations (\ref{Cn}) with even $n$
are equivalent
to the Klein-Gordon equation
let us introduce
the inverse zweibein $h^\nu_\pm$ and rewrite the system of equations
(\ref{Cn}) in the form
\be
\label{inv}
h^\nu_+ D_\nu \phi_n = \beta (n) \phi_{n-2}\,,\qquad
h^\nu_- D_\nu \phi_n = \alpha (n) \phi_{n+2}\,.
\ee
One observes that these equations
with $n=0$ express the fields $\phi_{\pm 2}$ in terms of
the first space-time derivatives of $\phi_{0}$.
Then the equations (\ref{inv}) with
$n=\pm 2 $
contain the Klein-Gordon equation and express the
fields $\phi_{\pm 4}$ via second space-time derivatives of
$\phi_{0}$. Note that although the Klein-Gordon equation appears
twice, i.e. both in the first of the equations (\ref{inv}) with $n=2$ and
in the second one with $n=-2$, an appropriate combination of these equations
vanishes identically due to the Bianchi identities of the original equations
(\ref{Cn}) so that, effectively, the Klein-Gordon equation appears only
once.
Finally, one finds that all higher $n$ equations
in the system (\ref{inv}) either express the fields $\phi_m$ with $m\neq 0$
via higher derivatives of $\phi_0$ or encode all Bianchi identities for these
expressions imposing no additional dynamical conditions
on the field $\phi_0$.

As a result, the system
(\ref{Cn}) with even $n$ turns out to be dynamically equivalent to the original
Klein-Gordon equation supplemented with some constraints which express all
higher
$\phi_n$ via higher space-time derivatives of the dynamical field $\phi_0$.
A situation with fermions ($n$ is odd) is analogous.

\section{Perturbative Analysis}
To analyze the equations (\ref{1f}) perturbatively one considers the fields of
the form ${\bf W}={\bf W}_{vac} + {\bf w}$ and ${\bf B}
={\bf B}_{vac} + {\bf b}$
where ${\bf w}$ and ${\bf b}$ denote perturbations.  Propagating matter fields
belong to the mutually conjugated components $|b\rangle$ and $\langle b|$ of
${\bf b}$.  The linearized equations (\ref{1f}) in the sector
of the matter fields $|b\rangle$ read \be
\label{le02} d|b\rangle + w^{gr} |b\rangle =N |w \rangle \,.
\ee This equation implies, first, that
$|w \rangle$ expresses via the matter fields $|b\rangle$ and, second, that it
imposes some differential equations on those components of the matter fields
which are not proportional to  $N$. Let us show that the latter
differential equations are just the equations for free matter
fields analyzed in the previous section.

The linearized gauge transformation (\ref{g1}) for the field
$|b\rangle$ takes the form
$\delta |b\rangle =  N |\xi \rangle +O({\bf b})$.
This implies that the field $|b\rangle$ contains some
Higgs part which can be gauged away and a
reminder which is to be shown to describe matter fields.

The standard Fock representation for
$|b\rangle$ is
$|b\rangle = b^l (\hat{u}_+ ,\hat{u}_-
) \Pi$.
Since $N=\frac{1}{4i}\{\hat{z}_+ ,\hat{z}_-\}$=
$\frac{1}{4i}\{(\hat{u}+\hat{v})_+ ,(\hat{u}+\hat{v})_-\}$, the Higgs - type
component of the transformation law for $|b\rangle$ allows one to get rid of
any
polynomial in $\hat{u}_+ \hat{u}_-$ in $b^l$. As a result one can chose a gauge
with respect to the transform (\ref{g1}) with
\be \label{hg} b^l (u_+ ,u_- ) =
b^l_+ (u_+ ) +b^l_- (u_- ) +b^l_0\,,\qquad  b^l_+ (0 ) =b^l_- (0 )=0\,.
\ee
Fields of this form cannot be compensated further by virtue of transformations
(\ref{g1}) and therefore can describe some dynamical degrees of freedom. By
expanding (\ref{hg}) in powers of $u_\pm$ one observes that the structure of
the
gauge fixed mater field $b^l$ (\ref{hg}) is just of the form one expects for d2
matter fields from (\ref{Cn}).  To work out explicit form of the field
equations
one has to substitute (\ref{hg}) into (\ref{le02}), decompose the
left-hand-side
of (\ref{le02})
into a part proportional to $N$ which is compensated by an appropriate choice
of
$|\omega\rangle$ and a part depending either only on $u_+$ or only on $u_-$ as
in (\ref{hg}) which will impose some equations on $b^l$.  Let us give the final
result for the field equations and the value of the field $|w\rangle$ =$ w^l
(\hat{u}_\pm ) \Pi$:
\begin{eqnarray}
\label{me}
D b^l &=&\frac{1}{4i}
h^{+} \left( (u_+)^2 (b^l_+ +b^l_0 ) +i u_+ \dot{b}^l_- (0) -4\ddot{b}^l_- (u_-
)+ \int_0^1 ds\,(3s+1)\ddot{b}^l_- (su_- ) \right)\nonumber\\
&+&\frac{1}{4i} h^{-} \left( (u_-)^2 (b^l_- +b^l_0 )
-i u_- \dot{b}^l_+ (0) -4\ddot{b}^l_+ (u_+ )+ \int_0^1 ds\,(3s+1)\ddot{b}^l_+
(su_+ ) \right)\,,
\end{eqnarray}
\begin{eqnarray}
\label{wl}
w^l (u_\pm )&=&-\omega b^l (u_\pm ) + \frac{i}{2}h^{+} \left(iu_+ \int^1_0 ds
\dot{b}^l_- (su_- ) -\int_0^1 ds\,(2s+1)\ddot{b}^l_- (su_- ) \right)\nonumber\\
&\phantom{=}&\phantom{\omega_0 b^l (\hat{u}_\pm )} + \frac{i}{2}h^{-}
\left(iu_-
\int^1_0 ds \dot{b}^l_+ (su_+ ) +\int_0^1 ds\,(2s+1)\ddot{b}^l_+ (su_+ )
\right)
\end{eqnarray}
where $\dot{f}(x)$=
${\partial\over \partial x} f(x)$.

One can check directly that the equation (\ref{me}) is formally
consistent thus corresponding to some particular
case of the equations (\ref{Cn}) with the coefficients of the form
(\ref{ab}).
By expanding the function $f^l$ into power series in either
$u_+$ or $u_-$
one finds that the coefficients indeed satisfy the condition (\ref{ab})
with $\mu =3/16$. This value is not occasional. It equals to the
value of the $sl_2$ Casimir operator for
the realization (\ref{vacw}).
There is a possibility to generalize
the proposed scheme to an arbitrary mass which we will discuss elsewhere
\cite{prep}.
Let us note that the parameter $\mu$
is measured here in units of the inverse radius
of the background AdS space-time and therefore tends to zero
in the flat limit.

Thus it is shown that the linearized equations for $|b\rangle$
describe properly linearized dynamics for d2 matter fields.
Analogously one can analyze the conjugate sector of $\langle b|$
to show that it describes conjugate matter fields.

An important property which
we do not prove explicitly here is that
all other components in ${\bf W}$ and ${\bf B}$ do not carry
their own degrees of freedom.
This can be shown for example with the aid of  the method
developed in \cite{un} where it was argued that any system of
covariant constantness equations for zero forms
cannot describe propagating modes when these zero forms
carry some finite-dimensional representations of the space-time
symmetry algebra which gives rise to the vacuum gravitational field.
Actually, in the model under consideration all components of the
zero form ${\bf B}$ contained in $B$ and $\langle B\rangle$
decompose into a sum of only
finite-dimensional representations of the AdS algebra under the adjoint action
of the generators (\ref{gen}).  We will come back to a more detailed
discussion of this point elsewhere \cite{prep}.

Thus, the matter fields contained in $|b\rangle$ and the conjugated
fields $\langle b|$ are the only propagating degrees of freedom in the
system. All other fields are either auxiliary or
 mediate interactions of
the matter fields.
In particular this is the case for the
gravitational field which corresponds to the sector of the $w$ fields quadratic
in $\hat{y}_\pm$ and for its HS generalizations corresponding to higher powers
in $\hat{y}_\pm$.

\section{Conclusions}

In conclusion let us summarize some
important properties of the proposed equations.

Due to the form of the product law (\ref{fp})
the matter fields contribute quadratically to the equations for the
gravitational field and its HS analogues
as expected from the matter sources for the gravitational field.
A concrete structure of the matter sources is involved
and will be given elsewhere.

In the linearized analysis it is shown that the Lorentz symmetry
keeps the standard form in the field equations, i.e.
Lorentz connection occurs only through the standard Lorentz
covariant derivative.
This is important and not completely trivial
property that can be shown to remain valid
in all orders in interactions \cite{prep}.

The proposed equations (\ref{1f}) have a form of some zero
curvature conditions and therefore can be integrated explicitly at least
locally
\be
{\bf W}(x)= {\bf g}^{-1}(x)d{\bf g}(x)\,,\qquad
{\bf B}(x)= {\bf g}^{-1}(x){\bf B_0}{\bf g}(x)\,,
\ee
where ${\bf g}(x)$ is an arbitrary $x$-dependent invertible element
of ${\cal A}$ while ${\bf B_0}$ is an arbitrary $x$-independent element
of ${\cal A}$.
To interpret properly this result one has to keep in mind that
there is an infinite collection of component fields in the model
due to the auxiliary variables $z_\pm$ and $y_\pm$
so that ${\bf B_0 }$ contains enough degrees of
freedom to describe all modes of a relativistic field
\cite{un}. Actually, it follows from the analysis of the
section 3 that one can analogously integrate the free mater equations
in terms of values of the set of fields $\phi_n$ in some point of
space-time (see also \cite{un}), which in their turn can be identified with
the fields $b^l_\pm (u_\pm ,x)$ (\ref{hg}). The
fact that the same phenomenon takes place for the non-linear model is due to
the specific HS interaction terms which make the system integrable.

\bigskip
\noindent
{\bf Acknowledgments}
\bigskip

I am very grateful to O.~Ogievetsky, A.~Tseytlin and I.~Tyutin
for useful discussions. The research described in this publication
was made possible
in part by Grant \# MQM000 from the International Science Foundation.
This work was supported in part by the Russian Basic Research
Foundation, grant
93-02-15541.

\end{document}